\documentclass[11pt]{JHEP3}
\renewcommand{\theequation}{\arabic{equation}}
\def\beq{\begin{equation}}
\def\eeq{\end{equation}}
\def\bea{\begin{eqnarray}}
\def\eea{\end{eqnarray}}

\def\nn{\nonumber}

\def\pa{\partial}

\newbox\pippobox

\def\bpsi{\bar{\psi}}
\def\bbeta{\bar{\beta}}

\def\ol{\overline}

\title{A Complete Solution of a Constrained System:\\
SUSY Monopole Quantum Mechanics}
\author{Soon-Tae Hong
\thanks{S.T.H. was supported by the Korea
Science and Engineering Foundation Grant R01-2000-00015.}
\\
        Department of Science Education, Ewha Womans University,
        Seoul 120-750 Korea
          E-mail: \email{soonhong@ewha.ac.kr}}
\author{Joohan Lee\\
        Department of Physics, University of Seoul,
        Seoul 130-743 Korea\\
        E-mail: \email{joohan@kerr.uos.ac.kr}}
\author{Tae Hoon Lee
\thanks{T.H.L. was supported by the Soongsil University
Research Fund.}
\\
        Department of Physics, Soongsil University,
        Seoul 156-743 Korea\\
        E-mail: \email{thlee@ssu.ac.kr}}
\author{Phillial Oh
\thanks{P.O. was supported by the Korea Science and
Engineering Foundation (KOSEF) grant through Science Research
Center (SRC) program of Center for Quantum Spacetime (CQUeST)
2005.}
\\        Department of Physics and Institute of Basic Science,
        Sungkyunkwan University, Suwon 440-746 Korea\\
        E-mail: \email{ploh@newton.skku.ac.kr}}

\abstract{We solve the quantum mechanical problem of a charged
particle on $S^2$ in the background of a magnetic monopole for both
bosonic and supersymmetric particles by constructing Hilbert space
and realizing the fundamental operators obeying complicated Dirac
bracket relations in terms of differential operators. We find the
complete energy eigenfunctions. Using the lowest energy eigenstates
we count the number of degeneracies and examine the supersymmetry
structure of the ground states in detail.}
\keywords{supersymmetric quantum mechanics; magnetic monopole;
operator ordering; supersymmetry breaking}

\preprint{hep-th/yymmnn}

\begin{document}

\section{Introduction}
\setcounter{equation}{0}
\renewcommand{\theequation}{\arabic{section}.\arabic{equation}}

The nonlinear sigma model and its supersymmetric generalization have
provided widespread applications in lower dimensional field theory
\cite{zakr}, string theory in curved space-time \cite{polc} and
supersymmetric quantum mechanics \cite{witt}, and their quantization
has been an important issue in theoretical physics. One of the
methods to approach the model is to use the constrained variables.
In such cases, it is well known that the canonical method has to be
replaced by the Dirac procedure \cite{henn}. However, it is very
hard, in general, to construct the Hilbert space and explicitly
realize the operators as differential operators due to highly
nonlinear nature of the Dirac approach. Since there does not exist a
general method to deal with such problems, each case has to be
treated separately. One of the purposes of this paper is to
demonstrate a complete quantization procedure for certain particular
supersymmetric nonlinear sigma models.

The system we are interested in is a supersymmetric quantum
mechanical particle moving on $S^2$ under the influence of Dirac
magnetic monopole~\cite{dira} located at the center. Since the work
of Wu and Yang \cite{wuya} revealing that a charged particle
interacting with a magnetic monopole can be described in terms of
the monopole harmonics and the system exhibits many interesting
features \cite{cole}, several supersymmetric models have been
proposed and various physical aspects of the supersymmetric
generalization have been investigated \cite{dhoke, dejong, clee,
others}. In particular, it was discovered \cite{dhoke} that the
system in $R^3$ admits $N=1$ supersymmetric generalization. It was
then found that the model has another hidden supersymmetry and
pointed out that this additional supersymmetry is related to the
model restricted to $S^2$ \cite{dejong}. $N=2$ supersymmetric model
on $S^2$ was studied in Ref. \cite{clee} and the complete energy
eigenfunctions were obtained. More recently, in Refs.
\cite{jlee,jlee2}, $N=2$ and $N=4$ supersymmetric models were
studied in detail using the $CP(1)$ model type of variables, and
certain issues related to the spontaneous supersymmetry breaking
were discussed. This choice of variables was adopted because it
allows relatively simple supersymmetric formulations of the model
and consequently $N=2$ and $N=4$ models could be treated in a
similar fashion. It also has the usual merits that the vector
potential for the magnetic monopole and consequently the Lagrangian
are free of singularity~\cite{bala} and one does not have to deal
with multi-valued action~\cite{wu}. However, in these references the
Hilbert space representation of the commutator algebra of the basic
quantum observables was lacking.

In this paper, we fill this gap by constructing the Hilbert space by
means of single valued functions on $S^3$ instead of on $S^2$ and
finding the differential operator representation of the quantum
observables. In the bosonic and $N=2$ cases, the solution to the
problem is well known in terms of the usual spherical variables.
However, in $N=4$ case the Hilbert space representation of the
quantum operators and the complete energy eigenfunctions had not
been given in the literature as far as we know. Furthermore, we show
that the complete energy eigenfunctions can be expressed in terms of
certain simple polynomials of $z_i$ and $\bar{z}_i$, $i=(1,2)$. This
result, as a byproduct, provides another way of writing the monopole
harmonics usually known in terms of the Jacobi polynomials. More
importantly, using the exact energy eigenfunctions we count the
number of ground state degeneracies and study the supersymmetric
structure of the ground state energy sector, and discuss the
important issue of spontaneous supersymmetry breaking in greater
detail. We also show that the Hamiltonian and the angular momentum
operators are related in such a way that generalizes the classical
results~\cite{cole}: The minimum angular momentum quantum number
$k_{\rm min}=|g|$ of a unit charged bosonic particle in the
background of the monopole of strength $g$ is replaced by $k_{\rm
min}=|g-\sigma|$ in the case of a supersymmetric particle, where
$\sigma$ is the total spin component of the particle along the
radial direction.

This paper is organized as follows. In Sec. II, we analyze the
bosonic case using the $CP(1)$ model approach~\cite{adda}. In Sec.
III, the analysis is extended to $N=2$ supersymmetric case, and in
Sec. IV to the $N=4$ case. The summary and discussions are given in
Sec. V.

\section{Bosonic particle}
\setcounter{equation}{0}
\renewcommand{\theequation}{\arabic{section}.\arabic{equation}}

We start with the bosonic case. Although the quantization in this
case is well known in terms of the usual variables \cite{wuya,
clee}, we here present a detailed quantization procedure because the
complete quantization using $CP(1)$ type of variables are not well
known. Furthermore, the following sections on the supersymmetric
particles will rest heavily on the results of this section. Also,
the operator ordering ambiguity arising in the quantization
procedure is carefully treated.

For notational convenience, we set the electric charge $q=-1$ and
the mass of the particle $m=1$. Although the particle is moving on
$S^2$ (of unit radius), we will work with $S^3$ (also of unit
radius), which is a principal $U(1)$ bundle over $S^2$. To be
concrete, let us describe $S^3$ by two complex functions $(z_1,
z_2)$ satisfying $\bar z \cdot z\equiv\sum_{i=1}^{2} \vert z_i
\vert^2\ = 1$. The $U(1)$ group action on $S^3$ is given by $z
\rightarrow {\rm e}^{i\Lambda}z$, and the base manifold is $S^2$.
The projection map is given by $x^a\equiv \bar{z}\sigma^az$
($a=1,2,3$), which satisfy $x^ax^a=1$, and $\sigma^a$ denote Pauli
matrices. From the fiber bundle point of view, dynamics of a
particle moving on $S^2$ is described by the action of the form
$A[z(t)]$ which is invariant under the local $U(1)$ transformation.
This way of writing the action in terms of $S^3$ coordinates instead
of $S^2$ coordinates has certain mathematical advantages. For
instance, the vector potential for the magnetic monopole has no
string singularity when regarded as a field on $S^3$.

We write the Lagrangian as \beq L= 2|\dot{z} -z({\bar z}\cdot
\dot{z})|^{2}+ig(\bar{z}\cdot\dot{z}-\dot{\ol{z}}\cdot z).
\label{cplag}\eeq The first term is the kinetic part. It is
invariant under the local $U(1)$ transformation and reduces to the
standard kinetic energy term when written in terms of the $S^2$
coordinates. The second term represents the interaction of the
particle with the magnetic monopole of strength $g$ located at the
center of $S^2$. Under the local $U(1)$ transformation, it changes
only by a total time derivative term and the corresponding action is
$U(1)$-invariant. In every respect, $U(1)$ plays the role of the
electromagnetic gauge group. The interaction term again reduces to
the familiar form (up to a gauge transformation) when expressed in
terms of $S^2$ coordinates.

Let $p$ and $\bar{p}$ denote the momenta conjugate to the fields $z$
and $\bar{z}$ respectively, \beq p=2\left(\dot{\ol{z}}
-(\dot{\ol{z}}\cdot z)\bar{z}\right)+ig\bar{z},~~~
\bar{p}=2\left(\dot{z} -z({\bar z}\cdot \dot{z})\right)-igz.\eeq Due
to the constraint $\bar z \cdot z=1$, the momenta should satisfy
\beq p\cdot z=ig, ~~~ \bar{z}\cdot\bar{p}=-ig.\eeq Using the
constraints, the Hamiltonian can be written as \beq H_c=2|\dot{z}
-({\bar z}\cdot \dot{z}) z|^{2}=\frac{1}{2}p_iA_{ij}\bar{p}_j,
\label{cpham} \eeq where $A_{ij}$, defined by \beq
A_{ij}\equiv\delta_{ij}-z_i\bar{z}_j, \label{projector}\eeq
satisfies \beq A_{ij}z_j=0, ~~~ \bar{z}_iA_{ij}=0, ~~~
A_{ij}A_{jk}=A_{ik}, ~~~ \bar{A}_{ij}=A_{ji}.\eeq The standard
Poisson brackets are \beq
\{z_{i},p_{j}\}=\{\bar{z}_{i},\bar{p}_{j}\}=\delta_{ij},\eeq with
the remaining brackets being zero. A simple analysis shows that the
constraints can be classified into the following two second class
constraints \beq C_{1}=\bar{z}\cdot
z-1,~~~C_{2}=\bar{z}\cdot\bar{p}+p\cdot z, \label{cpsecconst} \eeq
and one first class constraint \beq
C_{0}=-i(\bar{z}\cdot\bar{p}-p\cdot z)+2g, \label{cpfirconst}\eeq
generating the $U(1)$ transformation. Because of the second class
constraints we need to calculate the Dirac brackets according to the
formula \beq \{A,B\}_{D}=\{A,B\}-\{A,C_{a}\}\Theta^{ab}\{C_{b},B\},
\label{diracbra} \eeq where $\Theta^{ab}$ is the inverse matrix of
$\Theta_{ab}=\{C_{a},C_{b}\}$. The result can be summarized as \beq
\begin{array}{ll}
\vspace{0.2cm}
\{p_{i},z_{j}\}_D=-\delta_{ij}+\frac{1}{2}\bar{z}_{i}z_{j},
&\{p_{i},\bar{z}_{j}\}_D=\frac{1}{2}\bar{z}_{i}\bar{z}_{j},\\
\vspace{0.2cm}
\{p_{i},p_{j}\}_D=\frac{1}{2}(p_{i}\bar{z}_{j}-p_{j}\bar{z}_{i}),
&\{\bar{p}_{i},p_{j}\}_D=\frac{1}{2}(\bar{z}_{j}\bar{p}_{i}-z_{i}p_{j})..
\label{dbrackets} \end{array}\eeq

The canonical quantization proceeds by replacing the classical
variables by the corresponding quantum operators\footnote{In this
paper we denote the quantum operators and the Hermitian adjoint by
the same symbols as the corresponding classical quantities and the
complex conjugate. The distinction should be clear from the
context.}, imposing the commutation relations according to the Dirac
quantization rule, $\{A,B\}_{D}\rightarrow -i[A,B]$ and the complex
conjugation becoming the Hermitian adjoint. In this step, there
usually appears the notorious problem of operator ordering
ambiguity. In our case, however, the ordering is fixed as follows:
\beq
\begin{array}{ll}
\vspace{0.2cm}
\left[p_{i},z_{j}\right]=-i\delta_{ij}+\frac{i}{2}\bar{z}_{i}z_{j},
&\left[p_{i},\bar{z}_{j}\right]=\frac{i}{2}\bar{z}_{i}\bar{z}_{j},\\
\vspace{0.2cm}
\left[p_{i},p_{j}\right]=\frac{i}{2}(p_{i}\bar{z}_{j}-p_{j}\bar{z}_{i}),
&\left[\bar{p}_{i},p_{j}\right]=\frac{i}{2}(\bar{z}_{j}\bar{p}_{i}-z_{i}p_{j}).
\label{cpcomm}
\end{array}
\eeq The above brackets should be supplemented by their Hermitian
adjoints and all trivial commutation relations were omitted. Note
that the brackets in the first line have no operator ordering
ambiguity. In the second line, the ordering of the first bracket is
fixed by the anti-symmetry property, while the ordering in the
second bracket is fixed by requiring that the variables
$(z_{i},\bar{z}_{i},p_{i},\bar{p}_{i})$ commute with the second
class constraint, $C_{2}=0$. This choice of ordering appeared before
in the bosonic $CP(1)$ model \cite{han}. Next, we need to quantize
the constraints. Obviously, $C_1$ has no ordering ambiguity. It can
also be shown that $C_2$ is free from ambiguity if we demand it be
self-adjoint.  First class constraint $C_0$, however, suffers from
the ordering ambiguity. Therefore, the quantum Gauss law constraint
should be of the form \beq
\hat{C}_0\equiv-i(\bar{z}\cdot\bar{p}-p\cdot z)+\alpha_{G} +2g=0,
\label{qgc}\eeq where the real constant $\alpha_G$ denotes the
ordering parameter.

Before trying to find the Hilbert space representation of Eq.
(\ref{cpcomm}) it is useful to decompose $p_i$ into two parts by
introducing $U_{B}$ and $B_{i}$ as follows: \beq U_{B}\equiv
-i(\bar{z}\cdot \bar{p}-p\cdot z),~~~ B_{i}\equiv
p_{i}+\frac{i}{2}U_{B}\bar{z}_{i}. \label{Bi}\eeq $\bar{B}_{i}$ is
defined to be the Hermitian conjugate of $B_{i}$. The second class
constraints (\ref{cpsecconst}) become \beq \bar{z}\cdot z-1=0,~~~~
B\cdot z=0,~~~\bar{z}\cdot\bar{B}=0, \label{cpbzzb}\eeq and the
quantum Gauss law constraint (\ref{qgc}) can be written as \beq
U_{B}+\alpha_G+2g=0. \label{qgc2}\eeq The basic commutation
relations (\ref{cpcomm}) can be rewritten, in terms of
$(B_{i},\bar{B}_{i},U_{B},z_{i},\bar{z}_{i})$, as follows: \beq
\begin{array}{ll} \vspace{0.2cm}
\left[U_{B},z_{i}\right]=z_{i},
&\left[U_{B},\bar{B}_{i}\right]=\bar{B}_{i},\\
\vspace{0.2cm}\left[B_{i},z_{j}\right]=-iA_{ji},
&\left[B_{i},\bar{z}_{j}\right]=0,\\
\vspace{0.2cm}
\left[B_{i},B_{j}\right]=-i(\bar{z}_{i}B_{j}-\bar{z}_{j}B_{i}),
&\left[B_{i},\bar{B}_{j}\right]=-\left(U_{B}-\frac{1}{2}\right)A_{ji},
\end{array}
\label{cpcommuzb} \eeq where $A_{ji}$ was defined in Eq.
(\ref{projector}). Again, we omitted the Hermitian adjoint and
trivial relations. Next, we need to define the quantum Hamiltonian.
There is again an ordering ambiguity. However, since a different
choice of ordering in our model produces only a constant term upon
imposing the Gauss law constraint, it suffices to choose one. We
choose the following Hamiltonian: \bea H &=&\frac{1}{4}\left(
p_iA_{ij}\bar{p}_j+\bar{p}_jA_{ij}p_i\right)\nn\\
&=&\frac{1}{4}\left(B_k\bar{B}_k+\bar{B}_kB_k-1\right).\label{bham}\eea

The angular momentum operator $K_{a}$ is defined by \beq
K_{a}=\frac{i}{2}\left(\bar{z}\sigma_{a}\bar{B}-B\sigma_{a}z+2i\,\bar{z}\sigma_{a}z\right)
-\frac{1}{2}\left(U_{B}-\frac{3}{2}\right)\bar{z}\sigma_{a}z,
\label{cpkiop} \eeq which satisfies the following commutation
relations: \beq
\begin{array}{ll} \vspace{0.2cm}
\left[K_{a},z_{i}\right]=-\frac{1}{2}(\sigma_{a}z)_{i},
&\left[K_{a},\bar{z}_{i}\right]=\frac{1}{2}(\bar{z}\sigma_{a})_{i},\\
\vspace{0.2cm}
\left[K_{a},\bar{B}_i\right]=-\frac{1}{2}(\sigma_{a}\bar{B})_i,
&\left[K_{a},B_i\right]=\frac{1}{2}(B\sigma_{a})_i,\\
\vspace{0.2cm}
\left[K_{a},U_{B}\right]=0,
&\left[K_{a},K_{b}\right]=i\epsilon_{abc}K_{c}. \label{cpkicomm}
\end{array}
\eeq Its square turns out to be related to the Hamiltonian as \beq
K_aK_a = 2H +\frac{1}{4}\left(U_B-\frac{3}{2}\right)^2.
\label{ksquared}\eeq

To construct the Hilbert space, we first consider the functions of
the form: \beq f(z,\bar{z})=c_{{i_1\cdots i_m} j_1\cdots
j_n}z_{i_1}\cdots z_{i_m}\bar{z}_{j_1}\cdots\bar{z}_{j_n},
\label{cpfzbarz} \eeq where $(m,n)$ are non-negative integers. The
complex coefficients $c_{i_1\cdots i_n j_1\cdots j_m}$ are totally
symmetric with respect to the interchange of any two indices
belonging to the same index group. Furthermore, we choose them to
vanish when indices from different groups are
contracted.\footnote{This is because if such two indices have a
non-trivial trace the function can be reduced using the constraint
$\bar{z}\cdot z-1=0$. Thus, the above restriction on the
coefficients can be regarded as a kind of irreducibility condition.}
Such functions with a fixed pair of integers $(m,n)$ generate a
complex vector bundle over $S^2$ of $(m,n)$ type. The Hilbert space
is defined as the direct sum of all such complex vector bundles.
Hermitian inner product is given by \beq \langle f_1,f_2 \rangle
=\int f_1^*(z, \bar{z})f_2(z, \bar{z})d\mu, \label{product}\eeq
where \beq d\mu=\frac{1}{2\pi}\delta (\bar{z}\cdot
z-1)d\bar{z}_{1}dz_{1}d\bar{z}_{2}dz_{2}.\label{measure} \eeq With a
straightforward calculation it can be shown that this integral is
the usual integral on the base manifold times the integral over the
$U(1)$ fiber.

On this Hilbert space we represent $z_i$ and $\bar{z}_i$ as
multiplications and $B_i$, $\bar{B}_{i}$ as follows: \bea
B_i&=&-iA_{kj}\frac{\partial}{\partial
z_k}A_{ji}=-iA_{ki}\frac{\partial}{\partial z_k}+i\bar{z}_i,\nn\\
\bar{B}_{i}&=&-iA_{ij}\frac{\partial}{\partial
\bar{z}_k}A_{jk}=-iA_{ik}\frac{\partial}{\partial \bar{z}_k}.
\label{Birep}\eea It can be shown that $\bar{B}_{i}$ is the
Hermitian adjoint of $B_i$ with respect to the product
(\ref{product}), and that they satisfy the constraint
(\ref{cpbzzb}). A further calculation shows that they reproduce the
commutator algebra (\ref{cpcommuzb}) if we represent $U_B$ by \beq
U_B = z_k\frac{\partial}{\partial
z_k}-\bar{z}_k\frac{\partial}{\partial \bar{z}_k}+
\frac{3}{2}.\label{U0}\eeq Using this result, representation for
other composite quantities can be easily found. For instance, the
angular momentum can be represented as \beq
K_a=\frac{1}{2}\left(\bar{z}\sigma_a\frac{\partial}{\partial\bar{z}}
-\frac{\partial}{\partial z}\sigma_a z\right).\eeq

The physical states are those satisfying the Gauss law constraint
(\ref{qgc2}), which we write as \beq
\tilde{U}_B+2\tilde{g}=0,\label{qgc2.5}\eeq where \beq
\tilde{U}_B\equiv U_B-\frac{3}{2}, ~~~~~ 2\tilde{g}\equiv
\frac{3}{2}+\alpha_{G}+2g. \label{defgtilde} \eeq Therefore, they
are represented by the functions (\ref{cpfzbarz}) with $(m,n)$
satisfying \beq (m-n)+2\tilde{g}=0. \label{qgc3}\eeq Note that
$2\tilde{g}$ must be an integer. This implies that $\alpha_G$ should
be a half integer if $2g$ is an integer according to the Dirac
quantization condition of the monopole charge.

In order to obtain the energy spectrum we introduce the following
operators: \beq a=\epsilon_{ij}B_{i}\bar{z}_{j},~~~
\bar{a}=\epsilon_{ij}z_{j}\bar{B}_{i}, \label{defaab}\eeq in terms
which the Hamiltonian (\ref{bham}) can be written as \beq
H=\frac{1}{4}(a\bar{a}+\bar{a}a),\label{hamina} \eeq where \beq
\left[a,\bar{a}\right]=-\tilde{U}_B,~~~
\left[\tilde{U}_B,a\right]=-2a,~~~
\left[\tilde{U}_B,\bar{a}\right]=2\bar{a}. \label{cpabara} \eeq
Explicit differential operator representation of $a$ and $\bar{a}$
is obtained by inserting Eq. (\ref{Birep}) into Eq. (\ref{defaab}):
\beq a=-i\epsilon_{ij}\bar{z}_{j}\frac{\pa}{\pa z_{i}},~~~
\bar{a}=-i\epsilon_{ij}z_{j}\frac{\pa}{\pa \bar{z}_{i}},
\label{cpaepsilon} \eeq and the Hamiltonian can be written as \beq
H=\frac{1}{4}\left(-2\frac{\pa}{\pa z_{i}}\frac{\pa}{\pa
\bar{z}_{i}} +2z_{i}\frac{\pa}{\pa z_{i}}\bar{z}_{j}\frac{\pa}{\pa
\bar{z}_{j}} +\bar{z}_{i}\frac{\pa}{\pa \bar{z}_{i}}+
z_{i}\frac{\pa}{\pa z_{i}}\right). \label{hamrep}\eeq When applied
to the functions (\ref{cpfzbarz}), the first term involving the
second order derivatives vanishes due to the irreducibility property
we required on the wavefunctions and the remaining terms give the
following energy spectrum: \beq
E=\frac{1}{4}(2mn+m+n),\label{spectmn} \eeq with $m$ and $n$ related
to each other by Eq. (\ref{qgc3}). If $\tilde{g}\ge 0$, $m$ can be
any non-negative integer. So, we set $m=s$, $(s=0,1,2,\cdots)$ and
$n=s+2\tilde{g}$. If $\tilde{g}\le 0$, on the other hand, the roles
of $m$ and $n$ are interchanged and we set $n=s$, $m=s-2\tilde{g}$.
Using this notation, the energy spectrum can be written as \beq E=
\frac{1}{2}\Big(s^2+s(2|\tilde{g}|+1)+|\tilde{g}|\Big).
\label{bspectrum} \eeq

The lowest energy corresponds to $s=0$. When $\tilde{g}\ge 0$,
ground states are described by anti-holomorphic functions of degree
$2\tilde{g}$ because $s=0$ in that case implies $m=0$,
$n=2\tilde{g}$. When $\tilde{g}\le 0$, we find that ground states
are described by holomorphic functions of degree $2|\tilde{g}|$ .
The number of independent ground states can be evaluated by counting
the number of independent components of totally symmetric
coefficient tensors of degree $2|\tilde{g}|$. Since each index can
take two values there are $2|\tilde{g}|+1$ independent ground
states. Higher energy states can be similarly obtained.

It is also interesting to look for the relations between the
Hamiltonian and angular momentum squared. Using Eqs. (\ref{qgc2.5})
and (\ref{defgtilde}), we find that Eq. (\ref{ksquared}) becomes
\beq K_aK_a= 2H+\tilde{g}^2. \label{HKrelation}\eeq The eigenvalues
of the angular momentum squared can be evaluated from this equation
using the energy spectrum (\ref{bspectrum}) and, as expected, the
result turns out to be $k(k+1)$ with \beq k=k_{\rm min}+s,~~~ k_{\rm
min}=|\tilde{g}|.\eeq Note that if $|\tilde{g}|$ is a half integer
(or an integer), so must be $k$. In terms of this angular momentum
quantum number the energy spectrum can be written as \beq
E=\frac{1}{2}\left(k(k+1)-\tilde{g}^2\right).\label{egrelation}\eeq
The ground state energy $E_{\rm min}=\frac{1}{2}|\tilde{g}|$ is
achieved when the angular momentum quantum number takes the minimum
value $k=|\tilde{g}|$. Since there should be $2k+1$ degenerate
states for a given $k$, it follows that the ground state degeneracy
is due to the angular momentum degeneracy. The relation
(\ref{egrelation}) agrees with the well known result obtained using
the conventional approach~\cite{cole} except that the usual monopole
charge $g$ is replaced with $\tilde{g}$, which could be interpreted
as the effective monopole charge. The effect of the ordering
parameter $\alpha_G$ can be absorbed if we redefine the monopole
charge. In particular, $\tilde{g}$ is the same as $g$ if
$\alpha_G=-\frac{3}{2}$ is chosen.

\section{$N=2$ supersymmetric particle}
\setcounter{equation}{0}
\renewcommand{\theequation}{\arabic{section}.\arabic{equation}}

We next extend the previous analysis to $N=2$ supersymmetric case.
$N=2$ supersymmetric monopole Lagrangian \cite{jlee} is given by
\beq L= 2|\dot{z} -z({\bar z}\cdot \dot{z} )
|^{2}+\frac{i}{2}(\bpsi\cdot\dot{\psi}-\dot{\bpsi}\cdot\psi)
-\frac{i}{2}(\bar{z}\cdot\dot{z}-\dot{\ol{z}}\cdot z)\bpsi\cdot\psi
+ig(\bar{z}\cdot\dot{z}-\dot{\ol{z}}\cdot z-i\bpsi\cdot\psi),
\label{susylagcom} \eeq where in addition to the bosonic degrees of
freedom $z_i$ there are also anti-commuting fermionic degrees of
freedom denoted by $\psi_i$. As before, the dots between the symbols
mean contractions of the indices.

The momenta $p$ and $\bar{p}$ conjugate to $z$ and $\bar{z}$,
respectively, are given by \beq p=2\left(\dot{\ol{z}}
-(\dot{\ol{z}}\cdot
z)\bar{z}\right)-\frac{i}{2}(\bpsi\cdot\psi-2g)\bar{z},~~~
\bar{p}=2\left(\dot{z} -z({\bar z}\cdot
\dot{z})\right)+\frac{i}{2}(\bpsi\cdot\psi-2g)z, \eeq and the
Hamiltonian is given by \beq H_c=2|\dot{z} -z({\bar
z}\cdot\dot{z})|^{2} -g\bpsi\cdot\psi
=\frac{1}{2}p_iA_{ij}\bar{p}_j-g\bpsi\cdot\psi. \label{susyham} \eeq
Due to supersymmetries, the bosonic constraints $C_1$ and $C_2$ of
the previous section should be supplemented by two more fermionic
constraints. They are obtained \cite{jlee} by applying
supertransformations on $C_1$. Altogether, there are four second
class constraints \beq C_{1}=\bar{z}\cdot z-1,~~~C_{2}=p\cdot
z+\bar{z}\cdot\bar{p},~~~ C_{3}=\bar{z}\cdot\psi,~~~C_{4}=\bpsi\cdot
z, \label{secconst} \eeq and one first class constraint
corresponding to the Gauss law constraint \beq
C_{0}=-i(\bar{z}\cdot\bar{p}-p\cdot z)-\bpsi\cdot\psi+2g.
\label{firconst}\eeq

The Poisson brackets are defined as usual \beq
\{z_{i},p_{j}\}=\{\bar{z}_{i},\bar{p}_{j}\}=\delta_{ij},
~~~\{\bpsi_{i},\psi_{j}\}=-i\delta_{ij}, \eeq and the Dirac brackets
can be easily computed using the formula (\ref{diracbra}). The
commutation relations consistent with the resulting Dirac brackets
can be written as \beq
\begin{array}{ll}
\vspace{0.2cm}
\left[p_{i},z_{j}\right]=-i\delta_{ij}+\frac{i}{2}\bar{z}_{i}z_{j},
&\left[p_{i},\bar{z}_{j}\right]=\frac{i}{2}\bar{z}_{i}\bar{z}_{j},\\
\vspace{0.2cm}
\left[p_{i},p_{j}\right]=\frac{i}{2}(p_{i}\bar{z}_{j}-p_{j}\bar{z}_{i}),
&\left[\bar{p}_{i},p_{j}\right]=\frac{i}{2}(\bar{z}_{j}\bar{p}_{i}-z_{i}p_{j})
+\bpsi_{j}\psi_{i}-\alpha_F A_{ij},\\
\left[\bpsi_{i},\psi_{j}\right]=A_{ji},
&\left[p_{i},\bpsi_{j}\right]=i\bpsi_{i}\bar{z}_{j}. \label{comm}
\end{array}
\eeq The square bracket between two fermionic operators should be
interpreted as the anticommutator. Apart from the appearance of the
fermion operators, the basic structure remains the same as in the
bosonic case. However, there is a small difference worth mentioning.
In contrast to the bosonic case, the commutator $[\bar{p}_i , p_j]$
acquires a term quadratic in the fermion operators, which causes a
new ordering ambiguity. In fact, the requirement that the second
class constraints commute with all other operators does not fix the
ordering completely. We introduced a real parameter $\alpha_F$ in
the commutator between $p$ and $\bar{p}$ to reflect this new kind of
ordering ambiguity. As in the bosonic case, the Gauss law constraint
suffers from the ordering ambiguity and we write the quantum Gauss
law constraint as \beq \hat{C}_0=-i(\bar{z}\cdot\bar{p}-p\cdot
z)-\bpsi\cdot\psi+\alpha_G+2g=0.\label{n2gauss}\eeq

It turns out that the commutation relations (\ref{comm}) become
greatly simplified if we introduce the following variables: \beq
\begin{array}{ll}
\vspace{0.2cm} \beta=\epsilon_{ij}z_{j}\psi_{i},&
\bbeta=\epsilon_{ij}\bpsi_{i}\bar{z}_{j}, \\
\vspace{0.2cm} w_{i}=p_{i}-\frac{i}{2}\bar{z}_{i}\bpsi\psi,&
\bar{w}_{i}=\bar{p}_{i}+\frac{i}{2}z_{i}\bpsi\psi. \end{array} \eeq
This amounts to solving the fermionic constraints because the old
variables automatically satisfying the constraints $C_3$ and $C_4$
can be readily recovered by the formula \beq
\psi_{i}=\epsilon_{ij}\bar{z}_{j}\beta,~~~
\bpsi_{i}=\epsilon_{ij}\bbeta z_{j}.\eeq In terms of these variables
the commutation relations (\ref{comm}) can be written as \beq
\begin{array}{ll}
\vspace{0.2cm}
\left[w_{i},z_{j}\right]=-i\delta_{ij}+\frac{i}{2}\bar{z}_{i}z_{j},
&\left[w_{i},\bar{z}_{j}\right]=\frac{i}{2}\bar{z}_{i}\bar{z}_{j},\\
\vspace{0.2cm}
\left[w_{i},w_{j}\right]=-\frac{i}{2}(\bar{z}_{i}w_{j}-\bar{z}_{j}w_{i}),
&\left[\bar{w}_{i},w_{j}\right]=\frac{i}{2}(\bar{z}_{j}\bar{w}_{i}-z_{i}w_{j})
-\alpha_F A_{ij},\\
\vspace{0.2cm} \left[w_{i},\beta\right]=0,
&\left[w_{i},\bbeta\right]=0,\\
\vspace{0.2cm} \left[\beta,\bbeta\right]=1, & \label{commww}
\end{array}
\eeq and the constraints as \beq C_{1}=\bar{z}\cdot
z-1,~~~C_{2}=\bar{z}\cdot\bar{w}+w\cdot
z,~~~\hat{C}_0=-i(\bar{z}\cdot\bar{w}-w\cdot
z)-2\bbeta\cdot\beta+\alpha_G+2g.\eeq

Note that the bosonic and fermionic sectors completely decouple from
each other, and $\beta$ and $\bbeta$ play the role of annihilation
and creation operators in the fermionic sector. Note also that the
operators $(w, \bar{w})$ satisfy the same commutation relations as
$(p, \bar{p})$ in Eq. (\ref{cpcomm}) except the term involving
$\alpha_F$. Due to this similarity, we can almost repeat the
analysis of the bosonic case. Namely, we decompose $w_i$ into two
parts, \beq U_B=-i(\bar{z}\cdot \bar{w}-w\cdot
z),~~~B_{i}=w_{i}+\frac{i}{2}U_B\bar{z}_{i}. \eeq The second class
constraints are again given by Eq. (\ref{cpbzzb}), and the Gauss law
constraint $\hat{C}_0$ in Eq. (\ref{n2gauss}) becomes \beq
U_B-2\bbeta\beta+\alpha_G+2g=0. \label{sqgc}\eeq Most of the
commutation relations (\ref{cpcommuzb}) remain the same. The only
difference is that the last equation is modified to \beq
\left[B_{i},\bar{B}_{j}\right]=-\left(U_B-\alpha_F-\frac{1}{2}\right)A_{ji}.
\label{bbbcom}\eeq

The supercharges are given by \beq
Q=p\cdot\psi=\epsilon_{ij}B_{i}\bar{z}_{j}\beta= a\beta,~~~
\bar{Q}=\bpsi\cdot\bar{p}=\bbeta\epsilon_{ij}z_{j}\bar{B}_{i}=\bbeta\bar{a},\eeq
where $a$ and $\bar{a}$ defined as in Eq. (\ref{defaab}) satisfy the
same form of commutation relations \beq
\left[a,\bar{a}\right]=-\tilde{U}_{B},~~~
\left[\tilde{U}_B,a\right]=-2a,~~~
\left[\tilde{U}_B,\bar{a}\right]=2\bar{a},\label{aabcom} \eeq if we
define $\tilde{U}_B$ by \beq \tilde{U}_{B}\equiv
U_{B}-\alpha_F-\frac{3}{2}.\label{ubtilde} \eeq Note the difference
from Eq. (\ref{defgtilde}). Here, we have absorbed the ordering
parameter $\alpha_F$ into the definition of the bosonic $U(1)$
generator. We choose as our quantum Hamiltonian \beq
H=\frac{1}{2}\left[Q,\bar{Q}\right]
=\frac{1}{2}\left(a\bar{a}-\left[a,\bar{a}\right]\bbeta\cdot\beta\right).\label{n2ham}\eeq
It can be shown that this supersymmetric Hamiltonian agrees with the
classical expression (\ref{susyham}) up to an ordering term.

We now proceed to construct the Hilbert space representation. The
Hilbert space consists of column vectors of the form \beq f=
\left(\begin{array}{c} f_1\\
f_2\end{array}\right),\label{n2states}\eeq where each entry belongs
to the bosonic Hilbert space considered in the previous section. The
Hermitian product is trivially extended. On this Hilbert space the
fermionic operators $\beta$ and $\bbeta$ are
represented by the matrices \beq \beta=\left(\begin{array}{cc} 0 &0\\
1 &0
\end{array}
\right),~~~ \bbeta=\left(\begin{array}{cc} 0 &1\\ 0 &0
\end{array}
\right). \eeq Bosonic operators $B_i$ and $\bar{B}_i$ are
represented, as before, by Eq. (\ref{Birep}), and their commutator
compared with Eq. (\ref{bbbcom}) yields the following
identification: \beq \tilde{U}_{B}= z_k\frac{\partial}{\partial
z_k}-\bar{z}_k\frac{\partial}{\partial
\bar{z}_k}.\label{ubtilrep}\eeq  The $N=2$ supersymmetric
Hamiltonian is related to our bosonic Hamiltonian (\ref{hamina}) as
follows: \beq H= \frac{1}{4}\left(a\bar{a}+\bar{a}a\right)
-\frac{1}{2}\left[a,\bar{a}\right]\Sigma
=\frac{1}{4}\left(a\bar{a}+\bar{a}a\right)-\tilde{g}\Sigma+\frac{1}{4},
\label{n2hamins} \eeq where we define the spin operator \beq
\Sigma\equiv \bbeta\beta-\frac{1}{2},\label{uf}\eeq which has
$\sigma=\pm (1/2)$ as its eigenvalues. In matrix form, the
Hamiltonian can be written as
\beq H=\frac{1}{2}\left(\begin{array}{cc} \bar{a}a &0\\
0 &a\bar{a}
\end{array}
\right). \eeq We write the Gauss law constraint (\ref{sqgc})
as\footnote{It can be shown that $\tilde{U}_F\equiv -2\Sigma$ is the
fermionic $U(1)$ generator and $\tilde{U}_B+\tilde{U}_F$ is the
total $U(1)$ generator. Note that the definition of $\tilde{g}$ here
differs from that of Ref. \cite{jlee} by a shift of $1/2$.} \beq
\tilde{U}_B-2\Sigma+ 2\tilde{g}=0, \label{sqgc1}\eeq with \beq
2\tilde{g}\equiv \frac{1}{2}+\alpha_F+\alpha_G+2g.
\label{n2gtilde}\eeq Note that the definition of the effective
magnetic charge $\tilde{g}$ is different\footnote{To avoid confusion
with notations it is important to remember that we are using the
same symbol if their physical meaning is the same but their
definitions may be different depending on what kind of particle we
are considering. In general, we will not repeat writing the
definition if it is the same as the previous one.} from Eq.
(\ref{defgtilde}) in bosonic case. In terms of the eigenvalues Eq.
(\ref{sqgc1}) can be rewritten as \beq
(m-n)+2(\tilde{g}-\sigma)=0.\label{sqgc2}\eeq This implies that
$2\tilde{g}$ should be an integer, and if $2g$ is also an integer
that the sum of the two ordering parameters $\alpha_F+\alpha_G$
should be a half-integer.

To obtain the energy spectrum apply the Hamiltonian (\ref{n2hamins})
to the components of the column vector (\ref{n2states}). From Eqs.
(\ref{hamina}), (\ref{spectmn}) and (\ref{ubtilrep}), we find \beq
E=\frac{1}{4}(2mn+m+n)+\frac{1}{2}(m-n)\sigma, \eeq where
$\sigma=1/2$ for the upper component and $\sigma=-1/2$ for the lower
component. As in the previous section we set $m=s$,
$n=s+2(\tilde{g}-\sigma)$ if $\tilde{g}-\sigma\ge0$, and $n=s$,
$m=s-2(\tilde{g}-\sigma)$ if $\tilde{g}-\sigma\le0$, where
$s=0,1,2,\cdots$. Then the energy spectrum can be written as \beq E=
\frac{1}{2}\Big(s^2+s(2|\tilde{g}-\sigma|+1)+|\tilde{g}-\sigma|\Big)
-(\tilde{g}-\sigma)\sigma. \label{espectrum} \eeq This energy
spectrum can be written in simple form if we use the angular
momentum quantum number. For this purpose, define the angular
momentum operator $K_{a}$ by \beq
K_{a}=\frac{i}{2}\left(\bar{z}\sigma_{a}\bar{B}
-B\sigma_{a}z+2i\,\bar{z}\sigma_{a}z\right)
-\frac{1}{2}\left(U_{B}-\alpha_F-\frac{3}{2}\right)\bar{z}\sigma_{a}z,
\label{n2k} \eeq which satisfies the commutation relations
(\ref{cpkicomm}). Note that the parameter $\alpha_F$ appears because
of its presence in the commutation relations (\ref{commww}).
Calculation similar to Eq. (\ref{ksquared}) yields \bea K^{2}&=&2H
+\frac{1}{4}\left(\tilde{U}_{B}-2\Sigma-1\right)
\left(\tilde{U}_{B}-2\Sigma+1\right)\nn\\&=& 2H
+\left(\tilde{g}-\frac{1}{2}\right)\left(\tilde{g}+\frac{1}{2}\right).
\label{jsqared} \eea From this equation and the energy spectrum
(\ref{espectrum}) we find that the eigenvalue of the squared angular
momentum is $k(k+1)$ with \beq k=k_{\rm min}+s, ~~~~ k_{\rm min}=
|\tilde{g}-\sigma|. \eeq Conversely, the energy spectrum can be
written in terms of $k$ as \beq E=\frac{1}{2}\left(k(k+1)-
(\tilde{g}-\frac{1}{2})(\tilde{g}+\frac{1}{2})\right).\eeq

Using this result we easily find that zero energy is achieved by the
upper component if $\tilde{g}\ge\frac{1}{2}$ and by the lower
component if $\tilde{g}\le-\frac{1}{2}$. We list below some of the
few zero energy states: \beq
\begin{array}{c}
\vspace{0.6cm}
\cdots\\
\end{array}
~~~
\begin{array}{c}\vspace{0.2cm}
\left(\begin{array}{c} 0\\ {z}{z}\end{array}\right)\\
\tilde{g}=-\frac{3}{2}
\end{array}
~~~
\begin{array}{c}\vspace{0.2cm}
\left(\begin{array}{c} 0\\ {z}\end{array}\right)\\
\tilde{g}=-1
\end{array}
~~~
\begin{array}{c}\vspace{0.2cm}
\left(\begin{array}{c} 0\\ 1\end{array}\right)\\
\tilde{g}=-\frac{1}{2}
\end{array}
~~~
\begin{array}{c}\vspace{0.2cm}
\left(\begin{array}{c} 1\\ 0\end{array}\right)\\
\tilde{g}=\frac{1}{2}
\end{array}
~~~
\begin{array}{c}\vspace{0.2cm}
\left(\begin{array}{c} \bar{z}\\ 0\end{array}\right)\\
\tilde{g}=1
\end{array}
~~~
\begin{array}{c}\vspace{0.2cm}
\left(\begin{array}{c} \bar{z}\bar{z}\\ 0\end{array}\right)\\
\tilde{g}=\frac{3}{2}
\end{array}
~~~
\begin{array}{c} \vspace{0.6cm}
\cdots\\
\end{array}
, \eeq where we omitted all the indices, indicating only the
polynomial nature of the state on $z$ and $\bar{z}$. The number of
degeneracies for these states is $2|\tilde{g}-\sigma|+1=2k+1$.
Excited states can be similarly constructed.

The case with $\tilde{g}=0$ is somewhat special because the energy
spectrum is $E=\frac{1}{2}(s+1)^2$ for both upper and lower
components. This means that there is no state invariant under the
supersymmetry transformations. The minimum energy sector in this
case consists of two copies of $k=\frac{1}{2}$ states
\beq \left(\begin{array}{c} 0\\
\bar{z}\end{array}\right), ~~~~~~ \left(\begin{array}{c} z\\
0\end{array}\right),\eeq which are related to each other by the
supersymmetries. The supersymmetry is spontaneously broken for
$\tilde{g}=0$.

\section{$N=4$ supersymmetric particle}
\setcounter{equation}{0}
\renewcommand{\theequation}{\arabic{section}.\arabic{equation}}

The Lagrangian for $N=4$ superparticle moving on $S^2$ is given
\cite{jlee2} by \bea L&=& 2|\dot{z} -z({\bar z}\cdot \dot{z})|^{2}
+\frac{i}{2}(\bpsi_\alpha\cdot\dot{\psi}_\alpha-\dot{\bpsi}_\alpha\cdot\psi_\alpha)
-\frac{i}{2}(\bar{z}\cdot\dot{z}-\dot{\ol{z}}\cdot
z)\bpsi_\alpha\cdot\psi_\alpha\nonumber\\
&-& \frac{1}{2}(\bpsi_\alpha \cdot\psi_\alpha)^2+
ig(\bar{z}\cdot\dot{z}-\dot{\ol{z}}\cdot
z-i\bpsi_\alpha\cdot\psi_\alpha), \label{n4susylagcom}\eea where the
fermion field now carries an additional index $\alpha=(1,2)$. Note
that this Lagrangian differs from $N=2$ Lagrangian
(\ref{susylagcom}) by the presence of the quartic fermionic
interaction term which is essential for the existence of $N=4$
supersymmetry.

Canonical quantization of the system goes in parallel with that of
$N=2$ system. Additional fermion indices are treated in an obvious
manner. Momenta $p$ and $\bar{p}$ conjugate, respectively, to the
fields $z$ and $\bar{z}$ are \beq p=2\left(\dot{\ol{z}}
-(\dot{\ol{z}}\cdot
z)\bar{z}\right)-\frac{i}{2}(\bpsi_\alpha\cdot\psi_\alpha-2g)\bar{z},~~~
\bar{p}=2\left(\dot{z} -z({\bar z}\cdot
\dot{z})\right)+\frac{i}{2}(\bpsi_\alpha\cdot\psi_\alpha-2g)z. \eeq
The classical Hamiltonian is \beq H_c=2|\dot{z} -z({\bar
z}\cdot\dot{z})|^{2}+\frac{1}{2}(\bpsi_\alpha
\cdot\psi_\alpha)^2-g\bpsi_\alpha\cdot\psi_\alpha. \label{n4susyham}
\eeq Constraints are trivially extended. We have the following six
second class constraints \beq C_{1}=\bar{z}\cdot z-1,~~~C_{2}=p\cdot
z+\bar{z}\cdot\bar{p},~~~
C_{3\alpha}=\bar{z}\cdot\psi_\alpha,~~~C_{4\alpha}=\bpsi_\alpha\cdot
z, \label{n4secconst} \eeq and one first class constraint, \beq
C_{0}=-i(\bar{z}\cdot\bar{p}-p\cdot
z)-\bpsi_\alpha\cdot\psi_\alpha+2g. \label{n4firconst}\eeq

Starting with the Poisson bracket relations \beq
\{z_{i},p_{j}\}=\{\bar{z}_{i},\bar{p}_{j}\}=\delta_{ij},~~~
\{\bpsi_{i\alpha},\psi_{j\beta}\}=-i\delta_{ij}\delta_{\alpha\beta},
\eeq we quantize the system according the Dirac scheme to find the
following quantum commutation relations \beq
\begin{array}{ll}
\vspace{0.2cm}
\left[p_{i},z_{j}\right]=-i\delta_{ij}+\frac{i}{2}\bar{z}_{i}z_{j},
&\left[p_{i},\bar{z}_{j}\right]=\frac{i}{2}\bar{z}_{i}\bar{z}_{j},\\
\vspace{0.2cm}
\left[p_{i},p_{j}\right]=\frac{i}{2}(p_{i}\bar{z}_{j}-p_{j}\bar{z}_{i}),
&\left[\bar{p}_{i},p_{j}\right]=\frac{i}{2}(\bar{z}_{j}\bar{p}_{i}-z_{i}p_{j})
+\bpsi_{j\alpha}\psi_{i\alpha}-\alpha_F A_{ij},\\
\left[\bpsi_{i\alpha},\psi_{j\beta}\right]=\delta_{\alpha\beta}A_{ji},
&\left[p_{i},\bpsi_{j\alpha}\right]=i\bpsi_{i\alpha}\bar{z}_{j}.
\label{n4comm}
\end{array}\eeq
They form a straightforward generalization of Eq. (\ref{comm}).

We then solve the constraints $C_{3\alpha}$ and $C_{4\alpha}$, as
before, by introducing $\beta_{\alpha}$ and $\bbeta_{\alpha}$ as
\beq \beta_{\alpha}=\epsilon_{ij}{z}_{j}\psi_{i\alpha},~~~
\bbeta_{\alpha}=\epsilon_{ij}\bar{z}_{j}\bpsi_{i\alpha},\eeq and
define $w_{i}$ and $\bar{w}_{i}$ by, \beq
w_{i}=p_{i}-\frac{i}{2}\bar{z}_{i}\bpsi_{\alpha}\psi_{\alpha},~~~
\bar{w}_{i}=\bar{p}_{i}+\frac{i}{2}z_{i}\bpsi_{\alpha}\psi_{\alpha}.
\eeq  Bosonic part of the commutation relations remains the same as
Eq. (\ref{commww}) and the commutators invoving fermions become \beq
\left[w_{i},\beta_{\alpha}\right]=0,~~~
\left[w_{i},\bbeta_{\alpha}\right]=0, ~~~
\left[\bbeta_{\alpha},\beta_{\beta}\right]=\delta_{\alpha\beta}.\eeq
This shows that the fermion sector again decouples from the bosonic
one and the number of fermion annihilation and creation operators is
doubled.

We can proceed to introduce $U_B$ and $B_i$ and represent them on
the Hilbert space as in $N=2$ case. Because there are two fermion
creation operators the number of components of state vectors is
increased to four. Fermion operators $\bbeta_{\alpha}$ and
$\beta_{\alpha}$ can be represented by $4\times4$ matrices as
follows: \beq
\beta_{1}=\left(\begin{array}{cccc} 0 &0 &0 &0\\
1 &0 &0 &0\\
0 &0 &0 &0\\
0 &0 &1 &0
\end{array}
\right),~~~
\beta_{2}=\left(\begin{array}{cccc} 0 &0 &0 &0\\
0 &0 &0 &0\\
1 &0 &0 &0\\
0 &-1 &0 &0
\end{array}
\right),~~~
\bbeta_{1}=\left(\begin{array}{cccc} 0 &1 &0 &0\\
0 &0 &0 &0\\
0 &0 &0 &1\\
0 &0 &0 &0
\end{array}
\right),~~~
\bbeta_{2}=\left(\begin{array}{cccc} 0 &0 &1 &0\\
0 &0 &0 &-1\\
0 &0 &0 &0\\
0 &0 &0 &0
\end{array}
\right). \eeq The supercharges are given by \beq
Q_{\alpha}=p\cdot\psi_{\alpha}=\epsilon_{ij}B_{i}\bar{z}_{j}\beta_{\alpha}\equiv
a\beta_{\alpha},~~~
\bar{Q}_{\alpha}=\bpsi_{\alpha}\cdot\bar{p}=\bbeta_{\alpha}\epsilon_{ij}z_{j}\bar{B}_{i}
\equiv\bbeta_{\alpha}\bar{a},\eeq and satisfy the commutation
relation \beq \left[{Q}_{\alpha}, \bar{Q}_\beta\right]=
a\bar{a}\delta_{\alpha\beta}-\left[a,\bar{a}\right]\bbeta_\beta
\beta_\alpha.\label{qabqb}\eeq  Unlike $N=2$ case, the $N=4$
supersymmetric Hamiltonian cannot be obtained simply by taking the
trace of this equation because the result does not commute with the
supercharges. Nevertheless, it can be shown\footnote{This
hamiltonian differs from the one used in Ref. \cite{jlee2} by the
constant term $\frac{1}{2}\tilde{g}$. See the discussion in Section
V.} that the Hamiltonian can be defined as \beq
H=\frac{1}{4}\left[{Q}_{\alpha}, \bar{Q}_\alpha\right]-
\frac{1}{2}\tilde{g}\Sigma,\label{hqqb}\eeq where $\Sigma$ and
$\tilde{g}$ are defined by \beq \Sigma\equiv \bbeta_\alpha
\beta_\alpha-1, ~~~ 2\tilde{g}\equiv
2g+\alpha_G+\alpha_F-\frac{1}{2},\eeq which differ from the
corresponding equations (\ref{uf}) and (\ref{n2gtilde}) in $N=2$
case. Note that the $N=4$ spin operator $\Sigma$ has eigenvalues
$\sigma=(1,0,0,-1)$. The Gauss law constraint maintains the same
form as Eqs. (\ref{sqgc1}) and (\ref{sqgc2}). From Eq. (\ref{qabqb})
we get \bea \frac{1}{4}\left[{Q}_{\alpha},
\bar{Q}_\alpha\right]&=&\frac{1}{2}a\bar{a}
-\frac{1}{4}\left[a,\bar{a}\right]\bbeta_\alpha \beta_\alpha\nn\\
&=&\frac{1}{4}(a\bar{a}+\bar{a}a)-\frac{1}{4}\left[a,\bar{a}\right]\Sigma.\eea
Inserting this equation into Eq. (\ref{hqqb}) and using the Gauss
law we can write Hamiltonian in the following form: \beq H =
\frac{1}{4}(a\bar{a}+\bar{a}a)-
\tilde{g}\Sigma+\frac{1}{2}\Sigma^2.\label{n4hamspin}\eeq In matrix
form, it becomes \beq
H=\frac{1}{4}(a\bar{a}+\bar{a}a)+\left(\begin{array}{cccc} \frac{1}{2}-\tilde{g}&0 &0 &0\\
0 &0 &0 &0\\
0 &0 &0 &0\\
0 &0 &0 &\frac{1}{2}+\tilde{g}
\end{array}
\right).\eeq

The energy spectrum immediately follows from Eq. (\ref{n4hamspin})
\beq E=\frac{1}{4}(2mn+m+n)-\tilde{g}\sigma +\frac{1}{2}\sigma^2,
\eeq which in terms of the notation used in Eqs. (\ref{bspectrum})
and (\ref{espectrum}) can be written as \beq E=
\frac{1}{2}\Big(s^2+s(2|\tilde{g}-\sigma|+1)+|\tilde{g}-\sigma|\Big)
-\tilde{g}\sigma+\frac{1}{2}\sigma^2.\label{n4spectrum}\eeq It is
useful to express the energy spectrum in terms of the angular
momentum quantum number. It turns out that the angular momentum
operator $K_{a}$ in $N=4$ case has the same expression as Eq.
(\ref{n2k}) and the calculation of its square yields \beq
K^{2}=2H+\tilde{g}^2. \label{n4jsqared} \eeq Using Eq.
(\ref{n4spectrum}) we again find that the spectrum for $K^2$ is
$k(k+1)$, where \beq k=k_{\rm min}+s, ~~~~ k_{\rm min}=
|\tilde{g}-\sigma|. \eeq Conversely, the energy spectrum can be
written as \beq E=\frac{1}{2}\left(k(k+1)- \tilde{g}^2\right).\eeq

For a given $\tilde{g}$, $E_{\rm min}$ is determined by $k_{\rm
min}$. We tabulate $k_{\rm min}$, $E_{\rm min}$ and by which states
these values are achieved for each values of $\tilde{g}$. \beq
\begin{array}{lllll}
\vspace{0.2cm} &k_{\rm min}=|\tilde{g}|-1, &E_{\rm
min}=-\frac{1}{2}|\tilde{g}|,
&~~~\sigma=-1, &\mbox{for $\tilde{g}\le-1$,}\\
\vspace{0.2cm}&k_{\rm min}=\frac{1}{2}, &E_{\rm min}=\frac{1}{2},
&\left\{\begin{array}{l}\sigma=-1\\
\sigma=0\end{array}\right\},
&\mbox{for $\tilde{g}=-\frac{1}{2}$,}\\
\vspace{0.2cm}&k_{\rm min}=0, &E_{\rm min}=0, &~~~\sigma=0,
&\mbox{for $\tilde{g}=0$,}\\
\vspace{0.2cm}&k_{\rm min}=\frac{1}{2}, &E_{\rm min}=\frac{1}{2},
&\left\{\begin{array}{l}\sigma=0\\
\sigma=+1\end{array}\right\}, &\mbox{for $\tilde{g}=+\frac{1}{2}$,}\\
\vspace{0.2cm} &k_{\rm min}=|\tilde{g}|-1, &E_{\rm
min}=-\frac{1}{2}|\tilde{g}|, &~~~\sigma=+1, &\mbox{for
$\tilde{g}\ge +1$.} \end{array}\label{n4ssground}\eeq We list below
supersymmetric ground states for a few values of $\tilde{g}$: \beq
\begin{array}{c}
\vspace{0.6cm}
\cdots\\
\end{array}
~~~
\begin{array}{c}\vspace{0.2cm}
\left(\begin{array}{c} 0\\ 0\\ 0\\ zz\end{array}\right)\\
\tilde{g}=-2
\end{array}
~~~
\begin{array}{c}\vspace{0.2cm}
\left(\begin{array}{c} 0\\ 0\\ 0\\ z\end{array}\right)\\
\tilde{g}=-\frac{3}{2}
\end{array}
~~~
\begin{array}{c}\vspace{0.2cm}
\left(\begin{array}{c} 0\\ 0\\ 0\\ 1\end{array}\right)\\
\tilde{g}=-1
\end{array}
~~~
\begin{array}{c}\vspace{0.2cm}
\left(\begin{array}{c} 0\\ 0\\ 1\\ 0\end{array}\right)\\
\tilde{g}=0
\end{array}
~~~
\begin{array}{c}\vspace{0.2cm}
\left(\begin{array}{c} 0\\ 1\\ 0\\ 0\end{array}\right)\\
\tilde{g}=0
\end{array}
~~~
\begin{array}{c}\vspace{0.2cm}
\left(\begin{array}{c} 1\\ 0\\ 0\\ 0\end{array}\right)\\
\tilde{g}=1
\end{array}
~~~
\begin{array}{c}\vspace{0.2cm}
\left(\begin{array}{c} \bar{z}\\ 0\\ 0\\ 0\end{array}\right)\\
\tilde{g}=\frac{3}{2}
\end{array}
~~~
\begin{array}{c}\vspace{0.2cm}
\left(\begin{array}{c} \bar{z}\bar{z}\\ 0\\ 0\\ 0\end{array}\right)\\
\tilde{g}=2
\end{array}
~~~
\begin{array}{c} \vspace{0.6cm}
\cdots\\
\end{array}
\eeq For each values of $\tilde{g}$ there are $2k+1$ independent
states, again showing that the degeneracy is entirely due to the
angular momentum degeneracy. Consider $\tilde{g}=-2$ case for
instance. Since $s=0$ and $\sigma=-1$ for these states, the angular
momentum quantum number should be $k=|\tilde{g}-\sigma|=1$, and
$2k+1=3$ agrees with the number of independent states given by
symmetric combinations $z_1z_1$, $z_2z_2$ and $z_1z_2+z_2z_1$.

For $\tilde{g}=\pm\frac{1}{2}$ there does not exist any state which
is invariant under the full $N=4$ supersymmetry because the minimum
energy $E_{\rm min}=\frac{1}{2}$ is greater than
$-\frac{1}{2}\tilde{g}\Sigma$, the energy value supersymmetric
invariant states should have as can be seen from Eq. (\ref{hqqb}).
The ground states for these values are given by \beq
\begin{array}{c}\vspace{0.2cm}
\left(\begin{array}{c} 0\\ 0\\ z\\ 0\end{array}\right),
\end{array}~~~
\begin{array}{c}\vspace{0.2cm} \left(\begin{array}{c} 0\\
z\\ 0\\ 0\end{array}\right),
\end{array}~~~
\begin{array}{c}\vspace{0.2cm}
\left(\begin{array}{c} 0\\ 0\\ 0\\ \bar{z}\end{array}\right)
\end{array}~~~ \mbox{for $\tilde{g}=-\frac{1}{2}$,} \eeq and
\beq
\begin{array}{c}\vspace{0.2cm}
\left(\begin{array}{c} 0\\ 0\\ \bar{z}\\ 0\end{array}\right),
\end{array}~~~
\begin{array}{c}\vspace{0.2cm} \left(\begin{array}{c} 0\\
\bar{z}\\ 0\\ 0\end{array}\right),
\end{array}~~~
\begin{array}{c}\vspace{0.2cm}
\left(\begin{array}{c} z\\ 0\\ 0\\ 0\end{array}\right)
\end{array}~~~ \mbox{for $\tilde{g}=\frac{1}{2}$,} \eeq
consisting of three copies of $k=\frac{1}{2}$ states, six of them
altogether. They are related by supersymmetric transformations. For
$\tilde{g}=-1/2$, for instance, the first and the third states are
related by $Q_1$ and $\bar{Q}_1$ and the second and the third states
are related by $Q_2$ and $\bar{Q}_2$ as in the following diagrams:
\beq
\begin{array}{c}\vspace{0.2cm}
\left(\begin{array}{c} 0\\ 0\\ -iz_2\\ 0\end{array}\right)
\end{array}~~~
\begin{array}{c}\vspace{0.2cm}
\begin{array}{c} \bar{Q}_{1}\\ \longleftarrow\\
\longrightarrow\\
Q_{1}\end{array}
\end{array}~~~
\begin{array}{c}\vspace{0.2cm}
\left(\begin{array}{c} 0\\ 0\\ 0\\ \bar{z}_1\end{array}\right)
\end{array}~~~
\begin{array}{c}\vspace{0.2cm}
\begin{array}{c} \bar{Q}_{2}\\ \longrightarrow\\
\longleftarrow\\
Q_{2}\end{array}
\end{array}~~~
\begin{array}{c}\vspace{0.2cm} \left(\begin{array}{c} 0\\
iz_2\\ 0\\ 0\end{array}\right),
\end{array}~~~ \eeq
\beq
\begin{array}{c}\vspace{0.2cm}
\left(\begin{array}{c} 0\\ 0\\ iz_1\\ 0\end{array}\right)
\end{array}~~~
\begin{array}{c}\vspace{0.2cm}
\begin{array}{c} \bar{Q}_{1}\\ \longleftarrow\\
\longrightarrow\\
Q_{1}\end{array}
\end{array}~~~
\begin{array}{c}\vspace{0.2cm}
\left(\begin{array}{c} 0\\ 0\\ 0\\ \bar{z}_2\end{array}\right)
\end{array}~~~
\begin{array}{c}\vspace{0.2cm}
\begin{array}{c} \bar{Q}_{2}\\ \longrightarrow\\
\longleftarrow\\
Q_{2}\end{array}
\end{array}~~~
\begin{array}{c}\vspace{0.2cm} \left(\begin{array}{c} 0\\
-iz_1\\ 0\\ 0\end{array}\right).
\end{array}~~~ \eeq
Note that the states in the second column of the above diagram are
not invariant under any real supertransformations. On the other
hand, the states on the left are killed by $Q_2$ and $\bar{Q}_2$,
and the ones on the right are annihilated by $Q_1$ and $\bar{Q}_1$.
>From this fact we conclude that the space of the ground states for
$\tilde{g}=\pm 1/2$ consists of a two dimensional subspace
consisting of the states not invariant under any supersymmetries and
a four dimensional subspace consisting of the states invariant under
$N=2$ supersymmetry.

\section{Summary and discussions}
\setcounter{equation}{0}
\renewcommand{\theequation}{\arabic{section}.\arabic{equation}}
We have presented a complete solution to the quantum mechanical
problem of a charged particle moving on $S^2$ in the background of a
magnetic monopole at the center, starting with the simplest case of
a bosonic particle and extending the results to the supersymmetric
cases. In studying this model we have used $CP(1)$ type of
coordinates. This choice of coordinates has a certain advantage over
the conventional one. On the other hand, the use of redundant
coordinates produces more constraints which make the quantization
difficult. In principle, transition from the classical Dirac
brackets to the quantum commutation relations is not unique due to
the operator ordering ambiguity. Moreover, quantization of the
constraints can also involve ordering ambiguities. In this work we
have carefully retained all the possible ordering terms and found
certain quantization conditions they have to satisfy and eventually
shown that their effects can be absorbed by redefining the magnetic
charge. The quantum Hamiltonian may also have operator ordering
ambiguities. In our model, after using the quantum Gauss law
constraint the ordering ambiguity amounts to adding a constant term
linear in the magnetic charge. We have chosen the Hamiltonian in
such a way that the energy spectrum respects the symmetry under the
simultaneous flip of the magnetic field and the spin, which
certainly holds in the classical model. We have also required the
minimum energy to be zero when $\tilde{g}$ vanishes. This condition
further fixes $\tilde{g}$-independent constant term.

We have constructed the Hilbert space representation of the
fundamental quantum commutation relations, which was lacking in the
previous work of Refs. \cite{jlee, jlee2}. Using this representation
we have found the complete energy eigenfunctions. In particular, the
ground states were studied in detail. Explicit functional forms were
presented and the number of degeneracies were counted. For those
values of $\tilde{g}$ for which the ground states are invariant
under all supersymmetries we have shown that the number of
degeneracy is $2k_{\rm min}+1$. In Refs. \cite{jlee, jlee2} it was
noted that for certain values of $\tilde{g}$, i.e., $\tilde{g}=0$ in
$N=2$ case and $\tilde{g}=\pm\frac{1}{2}$ in $N=4$ case, the
supersymmetry is spontaneously broken. In this work we have further
investigated the ground state structure for these particular values
of $\tilde{g}$ and it was shown that the ground states consists of
two copies of $k=\frac{1}{2}$ states in $N=2$ case and three copies
of $k=\frac{1}{2}$ states in $N=4$ case. It was further shown, in
$N=4$ case, that one of them is not invariant under any
supersymmetry transformation and the remaining two are invariant
under a half of the supersymmetry transformations. Also in $N=4$
case, we notice from Eq. (\ref{n4ssground}) that the energy of the
supersymmetric ground states for $|\tilde{g}| \ge 1$ is negative,
$E_{\rm min} =-\frac{1}{2}|\tilde{g}|$, which was not possible in
the $N=2$ case. This is a special feature of the $N=4$ system due to
the last term of Eq. (\ref{hqqb}). A similar type of relation
appears also in Ref. \cite{Spector}. In our model, it can be roughly
explained by saying that the magnetic field and spin interaction
term in Eq. (\ref{n4hamspin}) becomes dominant over the spin-spin
interaction term for large values of $|\tilde{g}|$.


There are several aspects of this work which deserve further
studies. It seems possible to extend our analysis to any number of
supersymmetries beyond $N=4$ and it would be interesting to see how
the symmetry breaking pattern continues. Next, noting that in our
quantum mechanics model supersymmetry is spontaneously broken for
some special values of the effective monopole charge $\tilde g$, it
would be interesting to investigate the issue of spontaneous
supersymmetry breaking in the field theoretical extensions of our
model, paying attentions to the role of operator ordering ambiguity
and checking whether these particular values of $\tilde g$ have any
special meaning. Another topic is to consider the system on the
fuzzy (super) sphere \cite{hatsuda} and analyze whether some new
features of spontaneous supersymmetry breaking occur on the fuzzy
(super) sphere. It would be also worth investigating the BRST
extension~\cite{brst} of our supersymmetric monopole system in which
the ordering ambiguities could be further addressed.

\acknowledgments We would like to thank the Asia Pacific Center
for Theoretical Physics (APCTP) for the hospitality during our
visit.


\end{document}